\documentclass[12pt]{JHEP}
\usepackage{amsmath,latexsym,amssymb}

\newcommand{\beq}{\begin{equation}}
\newcommand{\bea}{\begin{eqnarray}}
\newcommand{\eea}{\end{eqnarray}}

\newcommand{\no}{\noindent}

\newcommand{\de}{\partial}

\title{The Fate of the Radion in Models with Metastable Graviton}
\author{L. Pilo \\ Scuola Normale Superiore, Pisa, Italy \\ INFN, Sezione di Pisa, Pisa Italy
\email{pilo@cibs.sns.it}}
\author{R. Rattazzi \\ Scuola Normale Superiore, Pisa, Italy \\ INFN, Sezione di Pisa, Pisa Italy
\email{rattazzi@cibs.sns.it}}
\author{ A. Zaffaroni \\ Scuola Normale Superiore, Pisa, Italy \\ INFN e Universit\`{a} di Milano-Bicocca, 
Milano, Italy
\email{alberto.zaffaroni@mi.infn.it}}
\abstract{
We clarify some general issues in models where gravity is localized at
intermediate distances. We introduce the radion mode, which is usually
neglected, and we point out that its role in the model is crucial.
We show that the brane bending effects discussed in the literature
can be obtained in a formalism where the physical origin
is manifest. The model violates
positivity of energy due to a negative tension brane, which induces a 
negative kinetic term for the radion. The very same effect that 
violates positivity
is responsible for the recovery of conventional Einstein gravity at
intermediate distances.} 
\preprint{ SNS-PH/00-08,  Bicocca-FT/00/03}
\begin{document}

\section{\large Introduction}
After the work of Randall and Sundrum (RS)
\cite{RS2} it is well understood that theories with non-compact 
extra-dimensions can localize gravity on three-branes \footnote{For an
earlier suggestion on graviton trapping see \cite{gog}.}.
A very interesting modification of this setup was recently proposed 
by Gregory, Rubakov and Sibiryakov (GRS) in \cite{GRS} (see also \cite{ross}).
The theory is effectively five-dimensional both at small and at large 
scales while it localizes gravity at intermediate scales. 
Since five-dimensional 
effects
are visible only at distances larger than the universe radius, the 
model can have an acceptable phenomenology
and many attractive features.
It was also noticed in \cite{witten,porrati1} that
the GRS model offers a completely new viewpoint on the cosmological 
constant problem.
Since the five-dimensional cosmological constant is zero, our 
brane-world will be naturally
flat. From a different perspective, 4-d gravity  disappears in our world at
 very large  distances possibly changing the impact of a non-zero 4-d vacuum 
energy.

It was unfortunately pointed out in \cite{witten} that the model has 
a serious drawback which makes it 
probably inconsistent. The  background indeed requires a 
negative-tension brane
in the bulk. Alternatively said, positivity of energy is violated.
Versions of the  same setup,
where the negative tension brane is smoothened in a kink solution of 
the five-dimensional
theory, suffer from a similar problem \cite{witten}. Such smooth 
solutions have a stress-energy tensor
that violates the weak-dominance energy condition. 
Many other papers \cite{csaki,porrati1,csaki1,rubakov1,porrati2,csaki2,kan} 
have addressed 
the question of consistency in the GRS model. 
%Putting aside the positivity issue,
These papers opened a debate on the capability of reproducing the 
correct Einstein theory
at intermediate distances. The claim that the GRS model has serious 
phenomenological
difficulties\cite{porrati1} has been criticized through an actual 
computation   
\cite{csaki1,rubakov1}. It was also pointed out that the very same
effect that is responsible for the agreement with the Einstein theory 
at intermediate distances,
shows up as a four-dimensional
anti-gravity force at very large distances \cite{rubakov1}.  
Based on an RG-inspired 
computation, in \cite{csaki2} it was
claimed that anti-gravity disappears at large distances
so that the GRS model is well behaved at all scales. The 
anti-gravity effect could be interpreted
as an effect of the internal inconsistence of the model \cite{porrati2},
therefore it  is extremely important 
to understand if it really exists.

The modest purpose of this letter is to reintroduce a mode which was 
mostly ignored in this debate and which
we believe should deserve much more credit. Indeed 
the position of the negative-tension brane is a modulus of the GRS solution
corresponding  to a localized
four-dimensional scalar. With some abuse of language
we will indicate this modulus as the radion. It was claimed by many authors
that the radion can be frozen or stabilized and therefore it is not 
crucial for the physics and internal consistency of the model.
However, 
as one can expect, and as
we will prove explicitly, the kinetic coefficient of the radion is
proportional to the brane tension, which is negative. 
This fact  makes the issue of radion stabilization 
very unclear and probably not even well
posed. We stress that the presence of a negative tension brane free to 
fluctuate in the bulk seems essential to the existence of quasi-localized
gravity.
The negative radion kinetic term very likely 
makes the theory inconsistent at some level, but we think it is worth
understanding more about this model. 

We show that an accurate 
analysis of the radion dynamics
sheds light on the various effects discussed in 
\cite{porrati1,csaki1,rubakov1,porrati2,csaki2}.
The radion is responsible  both for the recovery of Einstein gravity at 
intermediate distances and for the negative energy problem.
The violation of energy positivity 
does not happen at experimentally accessible scales, but
at very large distances and in the form of anti-gravity,  therefore 
the GRS model could be phenomenologically acceptable.
We stress that 
all these effects have already been discussed in  various papers
\cite{csaki,porrati1,csaki1,rubakov1,porrati2,csaki2}, but in a 
formalism that, in our opinion, somehow obscures
their simple and intuitive physical meaning. We prefer to choose a gauge 
for the gravity perturbations where there is no brane bending. In this gauge,
the propagating degrees of freedom have a direct interpretation as 
4-dimensional physical particles.

The role of the radion versus
the bending effect discussed in references \cite{csaki1,rubakov1} will
be clarified.  
Our results can be useful for studying other models as well. For
example, we discuss the non-compact model in \cite{LR}.
We also comment on the difference between our  results and 
reference \cite{csaki2}  which uses RG-inspired arguments.

\section{\large The Effective Action for Graviton and Radion}

%We briefly fix our notations and the models under discussion. 
We consider the five-dimensional background given by 
\beq ds^2=e^{2\phi (z)} \; \eta_{\mu \nu} \, dx^\mu dx^\nu  \, + \, dz^2 
\, ;
\label{e1}
\end{equation}
the flat 4-d metric $\eta_{\mu \nu}$ has the signature $(-1, 1,1,1)$, which
we use throughout the paper.
The metric (\ref{e1}) is the most general five-dimensional metric with 
four-dimensional Poincar\'e invariance and we mostly assume that it is a 
solution of a five-dimensional gravity theory with three-brane sources,
\beq 
\int dx^5 \sqrt{-\hat{g}} \, (2M^3 \,  R \, - \, \Lambda_a) \, - \, \sum_i 
\delta(z-z_i) \, \tau_i \, \int dx^4 \sqrt{-g_{{}_{in}}}
\label{e2} 
\end{equation}
where $g_{{}_{in}}$ is the induced metric on each brane.
The 5-d cosmological constant $\Lambda_a$ may vary in the different domains 
of space-time, labelled by $a$, delimited by three-branes. We shall
consider the case $\Lambda_a \leq 0$.
The derivative of the warp factor jumps at the three-brane
positions by an amount related to the tensions of the branes 
$\Delta\phi^\prime (z_c)=- \frac{\tau_i}{ 12 M^3}$. 
Tensions and cosmological constants in the bulk must satisfy these 
jump conditions. 
As a consequence there is a fine-tuning in order to get flat 
four-dimensional space. 

The non-compact RS model \cite{RS2} has $\phi (z)= -k|z|$. The bulk metric is 
AdS$_5$ with cosmological constant $\Lambda=-24M^3 k^2$ and 
there is a brane of positive tension $\tau_1=24M^3k$ at the origin. 
An orbifold $\mathbb{Z}_2$ symmetry $z\rightarrow -z$ is also imposed. The 
compact RS 
model \cite{RS} is obtained by compactifying the $z$ direction
on a circle and introducing a second brane at the  other orbifold 
point $z=r$ \cite{RS}. The tension of the second brane is negative
$\tau_2=-\tau_1$. Since the negative tension brane is sitting at an 
orbifold fixed point, the negative energy mode associated with
its fluctuations in the transverse direction is projected out. 

A modified non-compact RS model is obtained by introducing a second 
brane in the bulk at the
position $z=r$ \cite{LR}. From now on, this will be called the 
Lykken-Randall (LR) 
model\footnote{In \cite{LR}, the second brane is just a probe, with 
negligible
effects on the background. In this paper we consider the general case  of
a brane with arbitrary tension.}.
The five-dimensional cosmological constant is now different in the 
various regions of space.
The warp factor for the LR model is 
\begin{equation}
\phi(z) \, = \, \left[ \begin{array}{cc}
-k_{{}_{L}} z & 0 \leq z \leq r \\
-k_{{}_{R}} z \, + \, (k_{{}_{R}}-k_{{}_{L}})r  & z \geq r
\end{array} \right.   \quad .
\label{warpfact}
\end{equation}
So the 
cosmological constant equals respectively  $\Lambda_1=-24 M k_{{}_{L}}^2$ and 
$\Lambda_2=-24 M k_{{}_{R}}^2$ to the  left and to the right of the second brane.
The matching conditions at the
branes fix $\tau_1=24M^3k_{{}_{L}}, \; \tau_2=24M^3(k_{{}_{R}}-k_{{}_{L}})$. The right brane has 
positive tension only in the case $k_{{}_{R}}>k_{{}_{L}}$. Provided 
 $k_{{}_R}>0$,
the model localizes gravity and  a radion field. As shown below, in the
 LR model with  negative tension
brane the radion field has a negative kinetic term and a problem with 
positivity of energy. For models with
positive tension branes, the radion field can be
stabilized with a mechanism like that in \cite{goldwise}. In the 
compact RS and in the LR model (for observers living on the right brane),
phenomenological
agreement with observations requires that the radion is stabilized. 
Since we will mostly use the LR background as a toy model,
we will not stabilize the radion in this paper.

The GRS model can be formally obtained from the LR  by taking $k_{{}_{R}}=0$. 
The space is flat outside the brane region
and the theory is really five-dimensional. However, gravity is 
localized at intermediate scales \cite{GRS}. The four-dimensional
graviton exists as a metastable  state in the Kaluza-Klein (KK) 
spectrum \cite{GRS,csaki,porrati1}. The 
model can be conveniently thought of as a regularized
version of the RS model, with the position $r$ of the right brane as 
a regulator. When $r \rightarrow + \infty$ we recover the
RS model with a four-dimensional graviton at all finite scales. 
For finite $r$, the graviton is converted to a metastable state
decaying at large distances. Notice that 
$k_{{}_{R}}=0$ requires that the  brane at $x=r$  has negative tension. This
point is not invariant under the orbifold symmetry and
therefore the translational degrees of freedom of this brane 
are not projected out. The presence of a negative tension object free to 
fluctuate is expected to cause troubles at some point. 

%Normally one can make sense of a negative tension object provided it sits
%at an orbifold singularity, as in the compact 
%RS model \cite{RS}\footnote{Actually in the 
%compact RS model there remains a well to do radion corresponding to 
%the extra dimensional volume.}.
%In this case the orbifold projection freezes
%the dangerous translational mode. The GRS model, however, 
%needs two branes, one with positive
%and one with negative tension and it is crucial for the existence of 
%a metastable graviton
%that the positive tension one is sitting at the orbifold point. The 
%negative tension brane is  free to fluctuate in the bulk.
%This last feature seems to be essential to the existence of 
%quasi-localized gravity. 

The Lagrangian~(\ref{e2}) can be generalized to include scalar fields 
in the bulk. These extra scalar fields, with
a suitable potential, can be used to stabilize the radion. They can 
also be used to mimic the singular three-brane 
background with a smooth solution of a five-dimensional Einstein 
theory coupled to scalars. In this
language, a negative tension brane will appear as a kink solution of 
an ill-defined bulk theory. Positivity of
energy shows up in the smooth solution as the requirement that 
$\phi^{\prime\prime}\le 0$. This follows from
the weak dominance energy condition or, even more simply, from an 
equation of motion that
involves the kinetic term for the five-dimensional scalar fields 
$\lambda_a$, 
\beq
\phi^{\prime\prime}=-\sum_{ab} 
G^{ab}\partial\lambda_a\partial\lambda_b \label{e3}\quad . \end{equation}

Our purpose now is to study fluctuations of the background~(\ref{e1}) 
corresponding to the four-dimensional
graviton and radion.
Many results on the RS and GRS models have been obtained using a 
particular coordinate system where the fluctuations $h_{{\mu\nu}}$ are transverse 
and traceless. This simplifies the equations, but in this system the 
branes are bent. In order to read the physical potential on a brane,
the fluctuation is transformed to Gaussian Normal (GN)
coordinates with respect to the brane. The bending 
must be carefully taken into account in any computation of 
gravitational potentials, since it is crucially responsible for
the correct reproduction of the four-dimensional Einstein gravity 
\cite{garriga,lisa}. With two branes there are further complications,
simply because there is in general no coordinate system that is GN with respect
to both. In other words,
in coordinates where $\hat{g}_{zz}=1, \, \hat{g}_{\mu z}=0$ and the
brane at $z=0$ is flat, the brane at $z=r$ is in general bent. 
To avoid these kind of problems, we choose to work in a gauge where both 
branes are flat and where the 
physical interpretation of each mode is manifest. Our basic coordinate
frame is a slight generalization of GN to  a two brane system. This frame
can be easily constructed as follows.
We  start from GN coordinates with the respect to the brane in $z=0$; the
second brane will be defined by $F(x,z) =0$. For a small fluctuation
 $f(x)$,  $F=z-r-f(x)$, so that the bending of
the second brane is fully encoded in a 4-d scalar function.
 One can easily show (see the appendix) that by an infinitesimal coordinate
change $(x, \, z) \rightarrow (x^\prime, \, z^\prime)$ the metric can be put 
in the form
\beq
ds^{2} \, = \, a^{2}(z')\left [\eta_{\mu\nu} \, + \, \gamma_{\mu\nu}(x',z') 
\right]{dx'}^\mu {dx'}^\nu 
\, + \, (1\,+\chi(z')f(x') )\, {dz'}^2  \quad ;
\label{special}
\end{equation}
%\hat{g}_{AB} \, = \, \left( \begin{array}{cc} a^2(z^\prime) \left\{ \eta_{\mu \nu}
% \left[ 1 \, + \, 
%\chi(z^\prime) \, \phi(x^\prime) \right] \,
 %+ \,  
%\gamma_{\mu \nu}(x^\prime, \, z^\prime) \right \} & 0 \\
%0 & 1 \, + \,  \phi(x^\prime) \,  \chi(z^\prime) \end{array} \right) \quad ;
%\label{special}
%\end{equation} 
with the two branes sitting at $z^\prime = 0$ and $z^\prime = r$.
As a result, relaxing  the condition $\hat{g}_{zz}=1$ we can have ``parallel'' branes,
 still keeping  
$\hat{g}_{\mu z}=0$.  We shall omit the primes in what follows.
Following ref. \cite{rubrad}, in order to distinguish the fluctuations of
 spin 0 and spin 2, we further parametrize the linear perturbations as
\beq 
ds^2 \, = \, a(z,x)^2 \left[\eta_{\mu\nu} \, + \, \tilde{h}_{\mu\nu}(x,z) \, + \, 
2 \epsilon(z) \, 
\partial_\mu  
\partial_\nu f(x) \right]dx^\mu dx^\nu 
\, + \, b(z,x)^2 \, dz^2  \quad ;
\label{e4}
\end{equation}
where 
\bea
a^2& \, = \, & e^{-2kz} \left[1+B(z)f(x) \right] \nonumber\\
b^2 &=& \frac{(\partial_z \log a)^2}{ k^2} \simeq 
\left (1-\frac{B'}{k}f\right ) \; .
\label{e5}
\eea
The fields $f(x)$ and $\tilde h_{\mu\nu}(x,z)$ will end up
giving respectively the radion and the tower of spin 2 gravitons.
The case $\tilde{h}_{\mu \nu}(x,z) = h_{\mu\nu}(x)$ corresponds to  the 
four-dimensional fluctuations of the graviton.
% A similar form as been considered in \cite{rubrad}.
As usual, the wave-function of the graviton is proportional to the 
unperturbed warp factor $e^{-2kz}$ \cite{RS2}. 
Notice that the above
profile for $f(x)$ is the appropriate one for a modulus. When  $f$ is 
constant, the metric~(\ref{e4}) can be
reduced to the unperturbed AdS$_5$ with the trivial change of 
coordinates $\tilde z=z- \frac{B(z) f }{ 2k}$. With the parametrization
(\ref{e5}) and the additional relation
\beq B(z) \, = \, 2 \left[e^{2kz} \, + \, k \, e^{-2kz} \,\partial_z
\epsilon (z) \right] \; ,
\label{e5two}
\end{equation}
the equations of motion for $\tilde h_{\mu\nu}$ and $f$ decouple. 
This can be explicitly checked using the Einstein  equations:
the  $\mu\nu$ equation does not depend on $f$ so that
$\tilde h_{\mu\nu}$ solves the pure spin 2 equation. The other equations imply
$\Box f=0$.
\footnote{To find the effective lagrangian for the zero modes we could also
use the direct method of refs. \cite{csaki0,goldwise2}. Even though this 
parametrization cannot even solve the linearized Einstein equations 
\cite{rubrad} for non-trivial fields, it is adequate to describe the zero 
modes at the two derivative level. We explicitly checked this in the
compact RS model and in the LR model with $k_{{}_{R}}>0$. We preferred to use 
the method of ref. \cite{rubrad}, since it is completely well defined
also for $k_{{}_{R}}=0$ and because it nicely shows the physical
properties of the modes.}. From a 4-dimensional viewpoint
the choice in eq. \ref{e5two} corresponds
to a Weyl frame where the spin 2 fields are not kinetically mixed to 
the spin 0 radion.

Notice that the function $\epsilon(z)$ can be arbitrarily modified by
coordinate changes
that preserve $\hat{g}_{\mu z}=0$. However in  general these coordinate changes
also shift and bend the branes. This is why we kept $\epsilon$ free in the above.
As it will become clear below, the physically meaningful quantities are the values
of $\partial _z \epsilon$ at the brane positions. Fortunately these
quantities are not modified by reparametrizations that keep the branes straight
and are, moreover, fixed by the Israel junction conditions.
A particular metric of form (\ref{e5}) with $\epsilon\equiv 0$
was obtained in ref. \cite{rubrad} for the radion of the compact RS model.
Notice that if $B$ is of the form
\beq 
B(z) \, \sim  \, 2 \left[k \, e^{-2kz} \, \partial_z  \epsilon (z) \right] \; ,
\label{e7}
\end{equation}
the equations of motions are satisfied for any $f$, since the $f$ 
dependent part of the metric can be gauged away.
% There t and 
%the most general coordinate transformation that preserves $\hat{g}_{\mu 
%5}=0$. 
We stress that the great simplification obtained in \cite{rubrad} is that the 
equations of 
motion for $\tilde h_{\mu\nu}$ and $f$ are decoupled. 
Notice that in our gauge the calculation of the spin-two KK 
excitations goes through as usual.  The massive modes have an 
invertible 4-dimensional kinetic term and satisfy the 
condition $\tilde{h}^{(n) \mu }_\mu=0, \; \; 
\partial^\mu {{\tilde{h}}^{(n)}}_{\mu \nu}=0$ just by Einstein's equations.
For the zero mode
$\tilde h^{0}_{{\mu\nu}}(x,z)\equiv h_{\mu\nu}(x)$ we can use the remaining 4-d 
reparametrization $x^\mu\to x^\mu +\xi^{\mu}(x)$, $z\to z$ to go to a standard
gauge, for instance the harmonic one, without affecting the position 
of the branes. Then the propagator of our graviton zero mode (when it is 
normalizable)  has already the correct 4-dimensional tensor structure.

Our branes are located at $z=0$ and $z=r$, with a $\mathbb{Z}_2$ symmetry 
around the origin.
As discussed in \cite{rubrad} the radion corresponds to a non-trivial metric 
excitation which is however pure gauge in the region $|z|>r$.
So to find the radion, we need to patch together two
different copies of metric (\ref{e5}). In the region $|z|<r$, 
 $B$ will be given by eq. (\ref{e5two}) with 
$\epsilon=\epsilon_{{}_L}$, $k=k_{{}_L}$. For $|x|>r$,  $B$ is given by  
(see eq.  \ref{warpfact})
\beq 
B_{{}_R}(z) \, =  \, 2 \left[k \, e^{-2k_{{}_R}z + 2(k_{{}_R} - k_{{}_L})r} \, \partial_z  \epsilon_{{}_R} (z) 
\right] \quad .
\end{equation}
%$\epsilon = \epsilon_{{}_R}$, $k=k_{{}_{R}}$.
The matching conditions
between different regions in space-time with different values of $k$ 
requires continuity for the functions
$a,b, \epsilon,\partial_z \epsilon$. The conditions on the first 
derivative of $a$ is automatically satisfied with the ansatz for $a$ 
and $b$ in eq. (\ref{e5}). For instance, at $z=r$ the jump of the 
extrinsic curvature is  given by
\beq
\left[K_{\mu \nu} \right] \, = \, q a^2 \left[ \frac{1}{2} \, + \, 
\left(B \, + \, \frac{B^\prime}{2k} \right) \frac{1}{2} \right] 
\eta_{\mu \nu}
\, + \, \left[ a^2 q \epsilon \, + \, \left( \epsilon^\prime_{{}_R} 
\, - \, 
 \epsilon^\prime_{{}_L} \right)a^2 \right] \de_\mu \de_\nu f \; ,
\label{bip}
\end{equation}
with
$q=\lim_{z \rightarrow r} \left[\de_z \log a^2_{{}_R}-  
\de_z \log a^2_{{}_L} \right]$. 
As a result the Israel junction conditions,
\beq
\left [K_{\mu \nu} \right ] \, = \, - \left (2M^2 \right )^{-1} \, 
\left (T_{\mu \nu} \, - \, {1\over 3}
T_{\alpha \beta} \, g^{\alpha \beta}_{{}_{in}} \, {g_{{}_{in}}}_{\mu \nu} \right ) 
\label{jun2}
\end{equation} 
involving  the brane energy-momentum  tensor $T_{\mu \nu}= - {1\over 2} 
(\hat {g}_{zz})^{-1/2}
\, \tau_i \, {g_{{}_{in}}}_{\mu \nu}$, simply require that 
$ \epsilon^\prime$ is continuous through the brane.
   
The radion $f$ is a localized 4-dimensional field, whose kinetic term 
we want to calculate. Let us consider the LR model with generic 
$k_{{}_{L}}$ and $k_{{}_{R}}$.
Our general formulae can be applied to the
GRS model as well, provided the limit $k_{{}_{R}}\rightarrow 0$ exists. 
Set $\tilde{h}_{\mu\nu}= 0$ first.						
The matching conditions at the origin, imposing the $\mathbb{Z}_2$ symmetry, 
give $\partial_z\epsilon_{{}_L}(0)=0$, $B_{{}_{L}}(0)=2$. The matching conditions 
at $z=r$, $B_{{}_{L}}(r)=B_{{}_{R}}(r)$,
$\partial_z\epsilon_{{}_L} (r)=\partial_z\epsilon_{{}_R} (r)$
fix
\beq 
B_{{}_{L}}(r) \, = \, 2 \, e^{2k_{{}_{L}}r} \frac{k_{{}_{R}}}{  k_{{}_{R}}-k_{{}_{L}}}
\label{e8} \quad .
\end{equation}

Since we considered a pure gauge solution in the region outside the 
brane, the contribution of $|z|>r$ to the  action
is just a constant, which is irrelevant for our computation of the 
kinetic term.
The only contribution to the effective action comes from the region 
between the two branes. 
Expanding the Lagrangian (\ref{e2}) and considering only derivative
terms, we obtain
\bea 
{\cal L}&=&-M^3\int_{-r}^{r} dz \sqrt{-\hat{g}} \delta 
{\hat{g}}^{MN}(E_{MN})=2M^3\int_0^r dz \frac{3B^\prime (z)}
{ k_{{}_{L}}}(f\Box f)
\nonumber \\ 
&=& 6M^3 \, \frac{B_L(r)-B_L(0)}{ k_{{}_{L}}}(f\Box f) \, = \, \frac{24M^3}{ k_{{}_{L}}}
\left ( e^{2k_{{}_{L}}r} \frac{k_{{}_{R}}}{ k_{{}_{R}}-k_{{}_{L}}}-1\right )( \frac{f\Box f}{ 2})
\label{e9} \quad , 
\eea
where $E_{MN}$ are the five-dimensional equations of motion.
We always work on the full real axis for $z$ with a $\mathbb{Z}_2$ 
identification. The previous computation is greatly simplified
by the fact that the four-dimensional components $E_{\mu\nu}$ are 
identically zero\footnote{Since we considered
a flat four-dimensional metric, this is just the statement (or a 
check) 
that the equations of motion for $g$ and $f$ are decoupled.}. Notice 
that the final result, before integration in $z$,
is a total derivative. The kinetic term only depends on the value of 
the function $B(z)$ at the branes positions. This is a welcome fact since,
while the $B$ function itself is gauge dependent, its value on the branes
is fixed by the matching conditions.
%is invariant under coordinate changes that preserve the position
%of the branes (if the branes are displaced the ends of integration of 
%eq. \ref{e9} change as well).

Our 4-dimensional action truncated to the zero modes will be normalized as follows,
\beq 
{\cal L} \, = \, 2 \, \hat M_r^2\int dx^4 \sqrt{-g} \, R_4 \, + \, {C_r\over 2}\int 
\sqrt{-g} 
dx^4 \, f \Box f  \quad ,
\label{e11}
\end{equation}
with the kinetic term given by
\beq 
C_r \, = \, \frac{24M^3 }{ k_{{}_{L}}}\left (e^{2k_{{}_{L}} r}{k_{{}_{R}}\over k_{{}_{R}}-k_{{}_{L}}}-1
\right ) \quad .
\label{e12}
\end{equation}
The four-dimensional Planck mass is easily computed by integrating 
the graviton wave-function \cite{RS,RS2},
\beq 2\hat M_r^2 \, = \, 
4M^3\left ( \int_0^r e^{-2k_{{}_{L}} z}+\int_r^\infty 
e^{-2k_{{}_{R}} z +2(k_{{}_{R}}-k_{{}_{L}})r}\right ) \, = \, {2M^3\over k_{{}_{L}}}
\left ( 1+e^{-2k_{{}_{L}} r}{k_{{}_{L}}-k_{{}_{R}}\over k_{{}_{R}}}\right ) \; .
\label{e10}
\end{equation} 

Notice that $r$ explicitly appears in the above formulae. $r$
is related to the VEV of $f$ by a non-linear transformation.
\section{\large The Role of the Radion in the GRS Model} 
We can now extract information about the GRS model.
In the limit $k_{{}_{R}}\rightarrow 0$ the radion has a finite, and negative, 
kinetic term $C_r=-24M^3/k$, independent of $r$.
 Notice that the entire contribution to $C_r$ comes from $B(0)$, since
$B(r)=0$. As noticed in \cite{rubrad}, in the GRS
model the radion profile is zero on the second brane. This is  simply
due to the fact that the space is flat to the right.
The four-dimensional Planck mass, on the other hand, diverges in the 
limit $k_{{}_{R}}\rightarrow 0$. This is a signal
that the graviton zero mode is not normalizable in the GRS model. 
However, conventional four-dimensional gravity is mimicked
at intermediate scales by metastable states from the KK tower.
The integration over arbitrarily light KK modes produces a propagator 
for an effective  four-dimensional
graviton. To correctly reproduce the Einstein theory, this propagator 
must have the form
\beq 
<h_{\mu\nu} \, h_{\rho\sigma}> \, \sim \, \left ( 
{\eta_{\mu\rho} \, \eta _{\nu\sigma} \, + \, \eta_{\nu\rho} \, \eta_{\mu\sigma} \over 
2} \, - \, {\eta_{\mu\nu} \, \eta_{\rho\sigma} \over 2}\right )\frac{1}{q^2} \, = \, \left 
( {P_2\over 2} \, - \, {P_0\over 2}\right ){1\over q^2} \quad ,
\label{e13}
\end{equation}
where we have neglected terms involving $q_\mu$ in the tensor structure.
It was noticed in \cite{porrati1} that the KK modes unfortunately always 
give a contribution to the effective
propagator with a behavior appropriate to a 4-dimensional massive
graviton 
\beq 
{\Delta\over 4}({P_2\over 2} \, - \, {P_0\over 3})\frac{1}{q^2} \quad ,
\label{e14}
\end{equation}
where  $\Delta=k_{{}_{L}}/M^3$ was  explicitly computed in ref. \cite{GRS}.
Based 
on this discrepancy in the tensorial
structure of the propagator, the authors in \cite{porrati1} concluded 
that GRS-like models are in contradiction
with experimental results on gravitational forces. However physics is slightly
more subtle.
By five-dimensional general covariance, also the radion couples to the trace
of the energy momentum
%radion and KK states  always 
%couples to all the matter fields in the same way.  
so it contributes a factor $P_0/C_r$ to the effective 
propagator. 
Numerical factors felicitously combine in such a way that
 the full (KK + radion)  propagator will behave as prescribed by the 
four-dimensional Einstein theory.
Notice that the fact that the radion has a negative kinetic term is 
crucial for this argument.
It is certainly true that,
as stressed in \cite{porrati2}, 
whenever gravity is not localized but only obtained trough KK modes,
one needs a state with negative norm  to change  the $-1/3$ in the
propagator into $-1/2$\footnote{The authors of ref.\cite{porrati2} 
call generically this state a {\it ghost},
without trying to identify it in the model at hand.}. 
%We stress that 
%there is no
%need for {\it abstract} ghost states for explaining the behavior
%of gravity. The radion {\it must} be taken into account for a 
%consistent
%treatment of the problem and it is certainly a ghost, in the sense of 
%a field
%with negative kinetic term.}. 
In the GRS model, it not difficult to identify this state with the 
radion.

This result deserves some comments. That the discrepancy 
between massive and massless propagators can be
cured
was already pointed out in \cite{csaki1,rubakov1}, in analogy
with a similar phenomenon in the RS model.
A computation including the brane bending effect \cite{csaki1,
rubakov1} shows that the full graviton propagator at intermediate 
distances has the right form form ~(\ref{e13}). We just showed that the bending 
effect can be equivalently explained by the existence of a localized 
and physical radion mode.  At this point two remarks are in order.
First, notice that the equivalence is consistent because
 the bending computation in \cite{csaki1,rubakov1} is 
correct only for a radion that has not been stabilized\cite{garriga,tanaka}.
It has been shown (for the compact RS model)
\cite{tanaka} that the stabilization
mechanism induces an extra correction to the bending, as required by
physical intuition. Second, 
the analogy used by ref. \cite{csaki1} with the RS model is
somehow misleading. In the RS model, the bending effect is an artifact
of the gauge choice. By using our gauge where the graviton is not
transverse traceless, the 
four-dimensional
graviton automatically shows up with the right propagator. In the GRS 
model,
on the other hand, the bending has a physical meaning. This is due to 
the fact
that, in the GRS model, the graviton is not really a zero-mode, but
a metastable state made up with massive KK modes.

In conclusion,
 the behavior of gravity at intermediate scales in the GRS model is 
correct,
as already pointed out by many authors \cite{csaki1,rubakov1}.
At very large scales, the metastable graviton disappears and we are 
left with a radion with negative
kinetic term. Notice that on this point we disagree with 
ref. \cite{csaki2}.
They used a RG-inspired reasoning to argue that, even at very large 
distances, the sick negative energy
problem
%\footnote{Called in the various papers anti-gravity, bending 
%effect or ghost...} 
disappears and the model is well behaved at all scales. 
The argument in \cite{csaki2}  assumes implicitly that the radion is 
stabilized and explicitly that there is no
problem with the positivity of energy. Under these conditions, the physics
at large distances is apparently well behaved. The system of two branes would
look at very large scales like a single object without any 
trapped massless modes\footnote{This composite system has effectively zero 
tension. This, by itself, is not necessarily a problem, since
the composite brane sits at an orbifold fixed point.
Indeed one can consider a new model by flipping all the signs of
the brane tensions in GRS. At large distance we have a tensionless brane,
but with a radion of positive kinetic term trapped on it}.
However, the crucial issue in the GRS model is whether these 
conditions can be physically realized.
From the viewpoint of this paper the results of ref. \cite{csaki2} would
for instance be obtained if the radion had a mass of the order of 
$k_{{}_L}e^{-3k_{{}_L} r}$, the graviton width.
%\footnote{ Notice that the RG reasoning of 
%ref. \cite{csaki2} suggests a graviton width of the order of $k_{{}_L}e^{-k_{{}_L} r}$
%instead of $k_{{}_L}e^{-3k_{{}_L} r}$ as obtained by the explicit computation
%of refs. \cite{GRS,csaki1}.}. 
This way the graviton and radion potentials
would turn off together at large distances. However, giving a mass
to a field with negative kinetic term clearly does not remove the associated
instability. 
%It not clear how a field with negative kinetic term can be stabilized 
%and it
%is not even obvious if the question is well posed. 
%Moreover, even supposing that the radion can be stabilized, the
%computation in \cite{csaki2} is certainly correct
%at large distances, but now some extra effort is required to show 
%that the modification of the background due 
%to the stabilization mechanism does not affect the conclusion at 
%intermediate scale. 
%>From our effective lagrangian perspective a massive
%radion (whatever that means) would certainly modify
% gravity in the intermediate
%energy range. 
%This can also be seen in computations based on brane bending.
%In the compact RS model, for example, it has been 
%shown \cite{tanaka} that the stabilization
%mechanism induces an extra correction to the bending, as required by
%physical intuition. In practice, the bending effect, after 
%radion stabilization, is no longer simply determined by the warp factor in 
%the vicinity of the brane. This suggests that the bending effect
%of ref.\cite{csaki1} would no longer match the KK contribution to give the 
%right tensor structure of Einstein gravity. 
%The strong interplay between the metastable 
%graviton
%and the radion suggests that radion stabilization would 
%destroy the correct behavior of gravity at intermediate scales.  
%A very light (almost massless) radion and an extreme fine-tuning 
%is required to fulfill all
%conditions for a phenomenologically reasonable model. 
In any event, the 
negative energy problem will show up at some energy scale.

To further clarify the role of the radion and to check the consistency
of our method, we can use our results 
for models with $k_{{}_{R}}>0$ where we do have a graviton zero mode. 
The four-dimensional Lagrangian for the zero modes contains the graviton and the
radion. By construction, we do not have 
problems with gauge fixing and brane bending. The propagator for the 
graviton is the correct one, as
in equation~(\ref{e13}).
 Let us assume that the radion is not stabilized. 
Let us compare the bending computation with our method based on just 
the physical zero modes.
Without stabilization, the bending effect only depends on the warp 
factor derivative at the origin $z=0$, which is model independent. In 
particular,
it does not depend on $k_{{}_{R}}$. The four-dimensional Planck mass 
$\hat M_r$,  however,
depends on $k_{{}_{R}}$. The propagator on the $z=0$ brane truncated 
to zero modes is
\beq 
{1\over 8\hat M^2_r} \left (P_2 \, - \, {2\over 3}P_0 \right) \frac{1}{q^2} \, - \,  {1\over 
24M_L^2} \, \frac{P_0}{q^2} \quad ,
\label{fin1}
\end{equation}
where $M_L=M^3/k_{{}_L}$ is the  Planck mass in the RS model. Notice that
 Einstein gravity is not reproduced. We have instead a
tensor-scalar gravity, as expected from the existence of a radion 
field.  
From the viewpoint of our ``effective'' Lagrangian for zero modes, 
we have two contributions, ${1\over 8\hat M^2_r}(P_2-P_0)$ from the 
graviton,
and ${P_0\over C_r}$ from the radion. 
Agreement with the bending method requires 
\beq 
{1\over \hat M^2_r} \, - \,  {24\over C_r} \, = \, 
{1\over M_L^2} \, = \, {k_{{}_L}\over M^3} \quad .
\label{extra2}\end{equation}
Using eqs.~(\ref{e12}) and (\ref{e10}), we can easily check that this 
equation
is identically satisfied. This is another indication that our method is
equivalent to brane bending calculations.

One final comment concerning the role of the radion in the recovery
of locality is in order. Consider once more a LR model characterized by 
$k_{{}_{L}},k_{{}_{R}}$ and brane position $r$. We want to study the limit $r\to \infty$
with $k_{{}_{L}}$ fixed, as perceived by an observer doing experiments at a fixed 
distance on the Planck brane. This observer should recover the physics of the
non compact RS model with warp factor $k_{{}_{L}}$. The full propagator on this brane
has the form
\beq
\frac{1}{\hat M_r^2}(P_2 \, - \, P_0)\frac{1}{q^2} \, + \, \frac{1}{C_r}\frac{P_0}{q^2}
\, + \, \left (P_2 \, - \, \frac{2}{3}P_0 \right) \, G_{KK}(q^2) \quad ; 
\label{scaling}
\end{equation}
$G_{KK}$ indicates the contribution of the massive KK modes.
Now, if we keep $k_{{}_{R}}$ and $k_{{}_{L}}$ fixed and send $r\to \infty$ we have that
$C_r\to \infty$, so that the radion disappears,
while $\hat M_r^2$ goes to the value $\to M^3/k_{{}_{L}}$ proper for
the non-compact RS model. Then the above formula gives the right leading $1/r$ 
potential of RS without any adjustment from the KK tower.  On the other hand,
consider a limit in which $r\to \infty$ with $k_{{}_{L}}$ fixed but with a scaling
$k_{{}_{R}}$ such that $C_r$ stays fixed. By eq.~(\ref{e12}) 
this situation
requires $k_{{}_{R}}\sim e^{-2k_{{}_{L}}r}k_{{}_{L}}\ll k_{{}_{L}}$, so that the bulk brane must have
negative tension. It is important to remark that with this limiting procedure
$\hat M_r^2$ goes to a value different than $M^3/k_{{}_{L}}$. At first sight this 
seems to violate locality, as the second brane should be going out of
the sight of the Planck observer. To reestablish locality it is necessary
that the $G_{KK}$ term from the KK tower develops an effective pole at $q^2=0$.
This can fix the coefficient of the $P_2$ term, but it certainly
upsets the $P_0$ term. The latter can only be adjusted by the propagation of
an extra scalar, and the finite $1/C_r$ radion term is there precisely for 
this. Notice that the particular choice $k_{{}_{R}}=0$ in this discussion
is just the GRS model. We stress that the need for a non-decoupling massless
and ghost-like radion is {\it always} associated with the presence of
a brane of negative tension.

This ghost, as 
suggested by ref. \cite{porrati2}, serves the purpose of giving a smooth 
limit as the KK resonance mass (width) goes to zero. This smoothness is 
precisely 
the requirement of locality. Something vaguely similar can be done for the 
simple
Abelian Higgs model. In the unitary gauge the lagrangian is
\beq
L \, = \, \frac{1}{4e^2} F_{\mu\nu}F^{\mu\nu} \, - \, m^2 A_\mu A^\mu \, + \,  
A_\mu J^\mu
\end{equation}
where $J^\mu$ is the matter field contribution to the current, the analogue
of the brane energy momentum in our case. In the limit $m\to 0$ the photon 
keeps
having 3 helicity states and moreover its propagator becomes singular.
A smooth limit can be taken by adding an object with negative kinetic
term and with lagrangian
\beq
L \, = \, \frac{1}{2}(\partial \Phi)^2+\frac{\Phi}{m} \partial_\mu J^\mu.
\end{equation}
Diagrammatically, the propagation of $\Phi$ eliminates the singular 
$k_\mu k_\nu/m^2$ terms in the photon exchange and eliminates the third
unwanted helicity state. Formally this can be seen by adding the above two
 lagrangians and redefining $A_\mu\to A_\mu +\partial_\mu \Phi/m$.
Now, $\Phi$ only appears as a Lagrange multiplier term 
$\Phi \partial_\mu A^\mu$: integrating over $\Phi$ we get a gauge fixing
 condition $\partial_\mu A^\mu=0$. So we get for $m\to 0$ a massless
photon (2 helicities) with an invertible kinetic term (in the Landau gauge).
For finite $m$, the above simply corresponds to
Stueckelberg's Lagrangian (or B-field formalism) in  Landau gauge.

In the course of this discussion we only focussed on
what sees an observer on the Planck brane, whatever $k_{{}_{R}}$ is. 
Notice than from this point of view, as soon as $k_{{}_{L}} r$ is large enough
there is no real need to stabilize the radion in order to recover a good 
approximation to Einstein gravity. This happens both for positive and negative
bulk brane tension, and in particular for the GRS case of $k_{{}_{R}}=0$.
On the other hand, if we had been concerned with what sees an
observer on the bulk brane, then the recovery of Einstein gravity would
have required the stabilization of the radion. This is because the radion
couples a lot more strongly to this brane (at least for $k_{{}_{R}}\not = 0$).

%We could also
%have stabilized the radion. If $k_{{}_{L}}<k_{{}_{R}}$ all the branes have positive 
%tension and the radion can be frozen
%with a mechanism like that proposed in \cite{goldwise}. We now expect 
%to 
%recover the standard Einstein gravity. This is almost obvious in the
%effective Lagrangian approach. As noticed in \cite{tanaka},
%the stabilization mechanism add a contribution to the bending effect.
%In the compact RS model, it was proven that after stabilization the 
%standard
%Einstein gravity is recovered \cite{tanaka}. It should not be difficult
%to repeat the same computation in the LR model.
\section{\large Conclusions}
In this letter, we have clarified the role of the radion in the GRS 
model. 
We want to stress that, in our opinion, 
even before any actual computation of how gravity
behaves in the GRS model, one must face 
the basic objection made in \cite{witten}. 
Positivity of energy is a basic requirement.
In a model where it is violated, 
we expect to see dangerous effects at some point.
Similar objections have been made also in \cite{porrati2}.
The interplay between violation of energy-positivity 
and recovery of Einstein gravity is however intriguing.
As already stressed in \cite{porrati1,porrati2},
a negative energy state seems to be necessary for recovering the 
correct
Einstein gravity, and our analysis confirms this fact.
However, it is true that the
negative energy effect only appears at very large distances, bigger 
than the universe radius, with a proper choice of parameters.
It is a four-dimensional negative energy effect in a five-dimensional 
flat universe.
It is not clear how such an effect can be consistently 
incorporated in a sensible theory. 
%In any event, this sickness is certainly 
%unpleasant if we aim to
%address the cosmological constant problem using the GRS model.
A possible question now is whether consistent modifications of the 
GRS model 
or mechanisms for solving the radion problem exist.
We believe that, contrary to the claim in \cite{csaki2},
this problem has not yet been  
solved in the GRS model. Since the negative norm state played a 
crucial role
in our analysis, at first look it seems a difficult task to find 
a positive-energy model that, at the same time, reproduces the correct
gravity behavior.
However,
the intrinsic attractiveness of the model and the 
fact the Einstein gravity is correctly
reproduced at intermediate scales strongly calls for some way out.

\vskip 0.8truecm
\no
{\bf \large A. Linear perturbations}
\vskip 0.5truecm

\no
Let us start from the GN metric 
\beq
\hat{g}_{AB} \, = \, \left( \begin{array}{cc} g_{\mu \nu}(x, \, z) & 0  \\
0  & 1 \end{array} \right) \quad .
\end{equation} 
with the first brane sitting at $z=0$ and the second brane 
defined by the equation
\beq
 z=r+f(x)\quad.
\label{2b}
\end{equation}
%Being the hypersurface corresponding to second brane non-singular, one can solve (\ref{2b}) for $z$ getting
%the parametric equation $z = r + f(x)$, with $r$ constant. 
Let us consider the coordinate change 
\beq
z \, = \,  z^\prime  \, + f(x) \, \chi(z^\prime) \, , \qquad \chi(0) \, = \, 0 \, ,   \qquad \chi(r) \, = 
\, 1 \quad .
\end{equation}
In the coordinates $(x, \, z^\prime)$ the branes are located at $z^\prime = 0$ and  $z^\prime = r$ and
the metric has the form
\beq
\hat{g}_{AB}^\prime \, = \, \left( \begin{array}{cc} {g'}_{\mu \nu}(x, \, z^\prime) 
%\, + \, K_{\mu \nu} \,  \chi( z^\prime) \phi  
& \chi( z^\prime) \, \de_\mu f(x) \\
\chi( z^\prime) \, \de_\mu f(x) & [1 \, + \, f(x) \, \de_{z'} \chi( 
z^\prime) ]^{2}  \end{array} 
\right) \quad .
\label{ndia}
\end{equation} 
%Where the metric $ K_{\mu \nu}$ is the extrinsic curvature of the $z = $ const. surfaces with the background
%metric $ \hat{g}_{AB}$. 
This parametrization shows, as it is obvious, that in order to keep 
the branes parallel it suffices to introduce a scalar field $f(x)$ . With a further 
coordinate transformation $x^\mu = {x'}^\mu \, + \, \xi^\mu (x^\prime, 
z^\prime)$ we can eliminate the off-diagonal terms. For our purposes it 
suffices to consider an infinitesimal bending $f$ and work in the linearized approximation.
Then in eq.(\ref{ndia}) we have 
\beq
 g'_{\mu \nu } \, = \, a^{2}(z) \, \left[ \eta_{\mu \nu } \, + \, 
\gamma_{\mu \nu }(x, z) \right] \quad 
\end{equation}
where $\gamma_{\mu\nu}$ represents a small perturbation.
%if
%\beq
%%%%%%%%%%%%%%%%%%%%%%%%%%%%%%%%%%%%%%%%%%%%%%%%%%%%%%%%%%%%%%%%%%%%%%%%%%%%%%%%%%%%%%%%%%%%%%%%%%
%                                                                                                %
%   \chi  \, \de_\mu \phi \, + \, \xi^\nu  K_{\mu \nu} \, + \, \de_z \xi_\mu \, = \, 0 \quad .   %
%   \end{equation}                                                                               %
%   In practice, it is enough for us to consider the case                                        %
%                                                                                                %
%%%%%%%%%%%%%%%%%%%%%%%%%%%%%%%%%%%%%%%%%%%%%%%%%%%%%%%%%%%%%%%%%%%%%%%%%%%%%%%%%%%%%%%%%%%%%%%%%%
%with $\cal{B}_{\mu \nu }$ representing  a small perturbation. Working at linearized level, 
The required transformation is 
\beq
\xi^{\mu }\, = \, \psi(z^\prime) \, \eta^{\mu\nu}\de_\nu \phi \, , \qquad 
\mbox{with } \quad \chi (z')\, + \,
a^{2}(z')\de_{z'} \psi(z') \, = \, 0 \; .
\end{equation} 
As a result, we get a metric (\ref{special}) 
\beq
ds^{2} \, = \, a^{2}¥(z')\left [\eta_{\mu\nu} \, + \, \gamma'_{\mu\nu}(x',z') 
\right]{dx'}^\mu {dx'}^\nu 
\, + \, [1\,+2\partial_{z'} \chi(z')f(x') ]\, {dz'}^2  \quad .
\end{equation}
%%%%%%%%%%%%%%%%%%%%%%%%%%%%%%%%%%%%%%%%%%%%%%%%%%%%%%%%%%%%%%%%%%%%%%%%%%%%%%%%%%%%%%%%%%%%%%%%
%                                                                                              %
%                                                                                              %
%   \bea                                                                                       %
%   &&ds^2 \, = \, \left \{                                                                    %
%   \left[ b(z^\prime) \, + \, \chi(z^\prime) \,  \phi(x^\prime)                               %
%   \, \de_z b \right] \eta_{\mu \nu }                                                         %
%   \, + \, b(z^\prime) \, {\cal B}_{\mu \nu } \, + \, 2 \, \psi(z^\prime) \,                  %
%   \de_{\mu^\prime} \de_{\nu^\prime} \phi \right \} d{x^\prime}^\mu  d{x^\prime}^\nu \nb \\   %
%   && \qquad \qquad  + \,                                                                     %
%   \left[1 \, + \, 2 \, \phi \, \de_{z^\prime} \chi   \right] {dz^\prime}^2                   %
%   \quad .                                                                                    %
%   \eea                                                                                       %
%                                                                                              %
%%%%%%%%%%%%%%%%%%%%%%%%%%%%%%%%%%%%%%%%%%%%%%%%%%%%%%%%%%%%%%%%%%%%%%%%%%%%%%%%%%%%%%%%%%%%%%%%
\vskip .2in
\noindent

\acknowledgments
%{\bf Acknowledgments} \vskip .1in
%\noindent
We thank M. Mintchev and A. Strumia for helpful discussions.

\noindent
A. Z. is partially supported by the European Commission TMR program ERBFMRX-CT96-0045,
wherein he is associated to the University of Torino. 
L. P. and R. R. are partially supported by the EC under TMR contract ERBFMRX-CT96-0090.


\begin{thebibliography}{6666666666}

\bibitem{RS2} L. Randall and R. Sundrum, Phys. Rev. Lett. 83 (1999) 
4690, hep-th/9906064.
\bibitem{gog} M. Gogberashvili, hep-th/9812296.
\bibitem{GRS} R. Gregory, V. A. Rubakov and S. M. Sibiryakov, 
hep-th/0002072.
\bibitem{ross}I. Kogan, S. Mouslopoulus, A. Papazoglou, G. Ross and J. Santiago, hep-ph/9912552;
I. Kogan and G. Ross, hep-th/0003074.
\bibitem{witten} E. Witten, hep-ph/0002297.
\bibitem{porrati1} G. Dvali, G. Gabadadze and M. Porrati, 
hep-th/0002190.
\bibitem{csaki} C. Csaki, J. Erlich and T. J. Hollowood, 
hep-th/0002161.
\bibitem{csaki1} C. Csaki, J. Erlich and T. J. Hollowood, 
hep-th/0003020.
\bibitem{rubakov1} R. Gregory, V. A. Rubakov and S. M. Sibiryakov, 
hep-th/0003045.
\bibitem{porrati2}  G. Dvali, G. Gabadadze and M. Porrati, 
hep-th/0002190, hep-th/0003054.
\bibitem{csaki2} C. Csaki, J. Erlich, T. J. Hollowood and J. Terning, 
hep-th/0003076.
\bibitem{kan} G. Kang and Y.S. Myung, hep-th/0003162.
\bibitem{goldwise} W. D. Goldberger, M. B. Wise, Phys. Rev. D60 
(1999) 107505, hep-ph/9907218;
 Phys. Rev. Lett. 83 (1999) 4922, hep-ph/9907447.
\bibitem{csaki0} C. Csaki, M. Graesser, L. Randall, J. Terning,
hep-ph/9911406.
\bibitem{goldwise2}W. D. Goldberger, M. B. Wise, hep-ph/9911457.
\bibitem{LR} J. Lykken and L. Randall, hep-th/99080.
\bibitem{RS} L. Randall and R. Sundrum, Phys. Rev. Lett. 83 (1999) 
3370, hep-ph/9905221.
\bibitem{garriga} J. Garriga and T. Tanaka, hep-th/9911055.
\bibitem{lisa} S. B. Giddings, E. Katz and L. Randall, hep-th/0002091.
\bibitem{rubrad} C. Charmousis, R. Gregory and  V. A. Rubakov, 
hep-th/9912160.
\bibitem{tanaka} T. Tanaka and X. Montes, hep-th/0001092.

\end{thebibliography}
\end{document}